\documentclass[10pt,superscriptaddress,twocolumn,amsmath,amssymb,aps,prl,showpacs]{revtex4-1}
\usepackage{mathrsfs}
\usepackage{array}
\usepackage{graphicx}
\usepackage{dcolumn}
\usepackage{bm}
\usepackage[title,titletoc,header]{appendix}
\usepackage{amssymb}
\usepackage{amsmath}
\usepackage{mathrsfs}
\usepackage{amsmath}
\usepackage{subfigure}
\usepackage{float}
\usepackage[title,titletoc,header]{appendix}
\usepackage{graphicx}
\usepackage{color}
\usepackage{amssymb}
\usepackage{graphicx}
\usepackage{verbatim}

\begin{document}

\title{Universal three-body bound states in mixed dimensions beyond the Efimov paradigm}
\author{Pengfei Zhang}
\affiliation{Institute for Advanced Study, Tsinghua University, Beijing 100084, China}
\author{Zhenhua Yu}
\email{huazhenyu2000@gmail.com}
\affiliation{School of Physics and Astronomy and TianQin Research Center for Gravitational Physics, Sun Yat-Sen University, Zhuhai 519082, China}

\date{\today }

\begin{abstract}
The Efimov effect was first predicted for three particles interacting at an $s$-wave resonance in three dimensions. Subsequent study showed that the same effect can be realized by considering two-body and three-body interactions in mixed dimensions. In this work, we consider the three-body problem of two bosonic $A$ atoms interacting with another single $B$ atom in mixed dimensions: The $A$ atoms are confined in a space of dimension $d_A$ and the $B$ atom in a space of dimension $d_B$, and there is an interspecies $s$-wave interaction in a $d_{\rm int}$-co-dimensional space accessible to both species. We find that when the $s$-wave interaction is tuned on resonance, there emerge an infinite series of universal three-body bound states for $\{d_A,d_B,d_{\rm int}\}=\{2,2,0\}$ and $\{2,3,1\}$. Going beyond the Efimov paradigm, the binding energies of these states follow the scaling $\ln|E_n|\sim-s(n\pi-\theta)^2/4$ with the scaling factor $s$ being unity for the former case and $\sqrt{m_B(2m_A+m_B)}/(m_A+m_B)$ for the latter. We discuss how our mixed dimensional systems can be realized in current cold atom experiment and how the effects of these universal three-body bound states can be detected.
\end{abstract}

\maketitle

Dimensionality plays a crucial role in determining properties of physical systems. In three dimensions, three bosons with resonant two-body $s$-wave scattering can form an infinite series of three-body bound states. These states were first predicted by V.~Efimov in 1970 and found to have their binding energies $E_n$ following a peculiar geometric scaling $\ln|E_n|=-2\pi n/s_0$, with $s_0\approx 1.006$ a universal constant \cite{Efimov1970}. Recent experiments succeeded in realizing $s$-wave resonant scattering in ultra-cold atomic gases by the technique of Feshbach resonance \cite{Chin}, and launched an extensive investigation of the Efimov effect through measuring three-body recombination rate \cite{grimm_cesium, Jochim_3b, hulet, Gross1, OHara1, OHara2, gross, Berninger,Jin, Chin2014,Weidemuller2014, Chin2016}, atom-dimer inelastic collisions \cite{grimm_ad, jochim_ad} and radio-frequency spectroscopy \cite{jochim_radio, ueda_radio}. Contemporary theoretical investigations further deepened our understanding of the emergent universality of the Efimov three-body parameter in atomic gases \cite{Chin_a, Wang1, wang2, Schmidt, Naidon, Weidemuller2016}.

In two dimensions, the Efimov effect ceases to exist, and instead the super Efimov effect takes place \cite{Nishida2013, Nishida2014, Zinner2014, Gridnev2014, Gao2015}. Nishida \emph{et al.}~discovered that three two-dimensional fermions with resonant two-body $p$-wave scattering also form an infinite series of three-body bound states. These bound states are called ``super Efimov" since their binding energies follow a dramatic scaling $\ln(-\ln|E_n|)\sim3\pi n/4$. Signatures of the super Efimov effect have been predicted in the observables such as the atom loss rate \cite{Gao2015} and the time of flight and radio-frequency spectrum of atomic gases \cite{Zhang2017}. It remains an open question whether there exist infinite series of universal three-body bound states
other than the Efimov ones in three dimensions and the super Efimov one in two dimension. 

In this work, we consider three-body problems in mixed dimensions. Previously Nishida and Tan showed how the Efimov physics can be realized in mixed dimensions by considering two-body and three-body interactions \cite{Nishida2008,Nishida2009,Nishida2011}. Scattering in mixed dimensions with ultra-cold gases has also been probed experimentally \cite{Nishida2010}. In our setup of the mixed dimensions, two bosonic $A$ atoms are confined in a space of dimension $d_A$ and a single $B$ atom in a space of dimension $d_B$; there is inter-species two-body interaction which occurs in a $d_{\rm int}$-co-dimensional space that both the $A$ and $B$ atoms can access. When the two-body interaction is tuned at an $s$-wave resonance, we find that for two cases $\{d_A,d_B,d_{\rm int}\}=\{2,2,0\}$ and $\{2,3,1\}$, there emerge an infinite series of universal three-body bound states whose binding energies obey the scaling $\ln|E_n|\sim-s(n\pi-\theta)^2/4$. The scaling factor $s$ is unity for the former case, and equals $\sqrt{\gamma(2+\gamma)}/(1+\gamma)$ for the latter. Here $\theta$ is a three-body parameter and $\gamma=m_B/m_A$ is the mass ratio with $m_A$ ($m_B$) being the mass of the $A$ ($B$) atoms. These  universal three-body bound states in mixed dimensions go beyond the Efimov paradigm. 
We discuss ways to realize our mixed dimension setup in current cold atom experiments and how the effects of these universal three-body bound states can be detected.

\emph{Universal limit cycles.} We start with considering the situation that two bosonic $A$ atoms and a single $B$ atom are confined in two dimensions. There is an $s$-wave inter-species two-body interaction between an $A$ and $B$ atom, which occurs when both atoms are at the origin of the two dimensional plane. We employ a two-channel effective field theory to model the system, and the Lagrangian is given by
\begin{align}
L=&\sum_{j=A,B}\int d^2\boldsymbol\rho \psi^\dagger_j(\boldsymbol\rho)\left(i\partial_t+\frac{\nabla^2}{2 m_j}\right)\psi_j(\boldsymbol\rho)+D^\dagger\left(i\partial_t-\bar\nu\right)D\notag\\
-&\bar g\left[D^\dagger\psi_A(\mathbf{0})\psi_B(\mathbf{0})+\text{H.c.}\right]-\bar v_3\psi_A^\dagger(\mathbf{0}) D^\dagger D \psi_A(\mathbf{0}),\label{model}
\end{align}
for which the momentum cut-off is $\Lambda$. The dimer field $D$ is defined only at the plane origin $\boldsymbol{\rho}=\{x,y\}=0$ and it consists of a single $A$ and $B$ atom. The interaction breaks down the translational invariance and does not conserve momentum. 

\begin{figure}
	\includegraphics[width=0.47\textwidth]{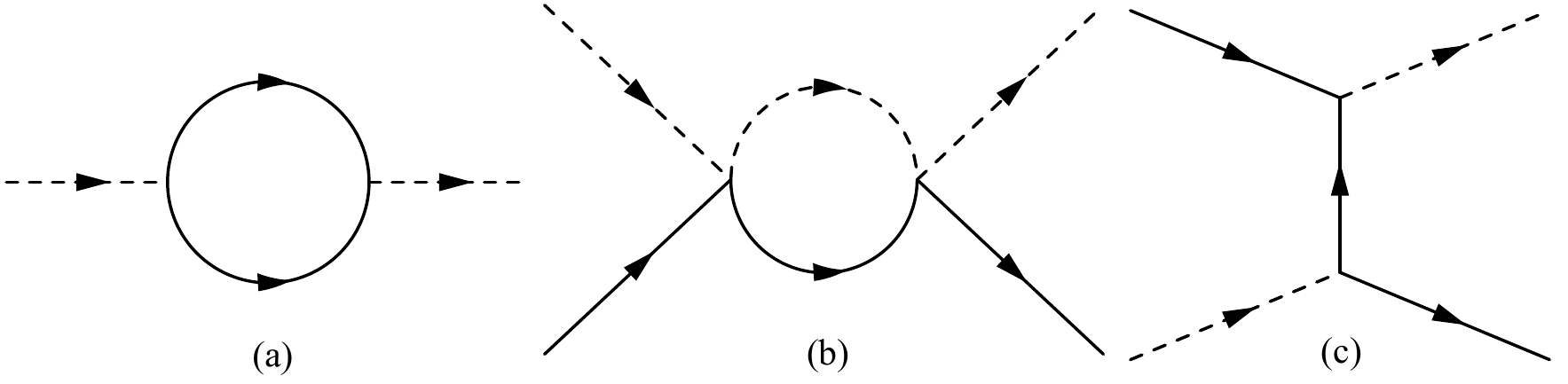}
	\caption{(a). The self-energy diagram for the dimer field $D$. (b-c). The second order diagrams for the atom-dimer scattering. The dashed lines represent the dimer field and the solid lines represent the atom fields.}
	\label{RG}
\end{figure}

The two-body scattering properties of Eq.~(\ref{model}) can be derived from the full propagator of the dimer field. We obtain the full propagator by summing up the bubble diagrams shown in FIG. \ref{RG} (a). This calculation yields the low energy expansion of the inter-species two-body scattering phase shift $\cot\delta(k)=-1/Sk^2+2\log(k R)/\pi$ with $k=\sqrt{m_A E}$ and $E$ being the total energy. The low energy scattering parameters, the scattering area $S$ and the effective range $R$, are related to the model parameters, $\bar\nu$, $\bar g$ and $\Lambda$, via the renormalization
\begin{align}
-\frac{ 1}{4\pi S}=&\frac{\bar\nu}{m_B \bar g^2}+\frac{f(\gamma)}{4\pi^2}\Lambda^2\label{s},\\
-2\ln(\Lambda R)=&\frac{4\pi^2}{m_Am_B\bar g^2}+1+f(1/\gamma)\label{r},
\end{align}
with $f(\gamma)=\ln(\gamma)/(1-\gamma)$. The effective range $R$ sets a natural scale for the short distance cutoff such that to use the model (\ref{model}) to describe low energy scattering, we take the hierarchy $k\ll\Lambda\ll1/R$ \cite{Zhang2017}. 

The low energy expansion of the phase shift $\cot\delta(k)$ can be understood by a corresponding single channel model: an $A$ atom and $B$ atom interact via a potential of range $r_0$ centering at the origin. Thus when both atoms are well away from the origin, the two-body wave-function satifies
\begin{align}
\left[-\frac{\nabla_A^2}{2m_A}-\frac{\nabla_B^2}{2m_B}\right]\Psi(\boldsymbol \rho_A, \boldsymbol \rho_B)=E\Psi(\boldsymbol \rho_A, \boldsymbol \rho_B).\label{h2}
\end{align}
We rescale the coordinates $\boldsymbol\rho_j=(x_j,y_j)$ and introduce 
\begin{align}
\mathbf{r}=\left\{x_A,y_{A},\sqrt{\gamma}x_{B},\sqrt{\gamma}y_{B}\right\}; \label{rrs}
\end{align}
in terms of the new coordinates $\mathbf r$, we recast Eq.~(\ref{h2}) into a Sch\"odinger equation for a single particle of mass $m_A$ in a four dimensional space. The wave-function for the $s$-wave scattering state takes the general form 
$\Psi(\mathbf r)=[\cot\delta(k) J_1(k r)- Y_1(k r)]/r$ with $J_1$ and $Y_1$ the Bessel functions. The phase shift $\cot\delta(k)$ is determined by requiring $\Psi(\mathbf r)$ satisfying a boundary condition at $r\gtrsim r_0$ \cite{Landau}. One can show that $\cot\delta(k)$ has the same low energy expansion structure as given above by the two-channel model (\ref{model}).

We study the three-body problem of two $A$ atoms and a single $B$ atom via the renormalization of $\bar v_3$. We fine tune $\bar\nu$ such that the two-body scattering is on resonance, i.e., $1/S=0$. The renormalization relation (\ref{r}) indicates that when one evolves the cutoff $\Lambda$ to the low energy limit, i.e., $\Lambda R\to0^+$,
\begin{align}
\frac{1}{\bar g^2}=-\frac{m_Am_B\ln(\Lambda R)}{2\pi^2},\label{g}
\end{align} 
and the model (\ref{model}) is in the weak coupling regime ($m_Am_B\bar g^2\ll1$). Thus we proceed to calculate the renormalization group (RG) equation of the three-body parameter $v_3$ perturbatively. The second order diagrams for the atom-dimer scattering are given by FIG. \ref{RG} (b) and (c). Note that due to the absence of momentum conservation, the diagram in FIG. \ref{RG} (c) involves a momentum integration although it looks at the tree level. We define  $\tilde v_3\equiv \bar v_3/m_B\bar g^2$ and $b\equiv-\ln \Lambda$, and find the RG equation to second order
\begin{align}
\frac{d \tilde v_3}{db}=-\frac{2\pi \tilde v_3^2}{b}-\frac{1}{\pi} \label{v3rg}.
\end{align}
In the low-energy limit, i.e., $\Lambda\to0$ and $1/b\to0^+$, we solve Eq.~\eqref{v3rg} by an ansatz $\tilde v_3=\sqrt{b}f(\sqrt{b})$ and neglecting corrections of order $1/b$; the solution is 
\begin{align}
\tilde v_3=\frac{\sqrt{b}}{\sqrt{2}\pi}\tan\left(-2\sqrt{2b}-\theta\right),\label{v3}
\end{align}
where $\theta$ is a three-body parameter determined by the short-range details of three-body interaction \cite{Braaten2006}. Equation (\ref{v3}) shows that $\tilde v_3$ diverges wherever $2\sqrt{2b}+\theta=n \pi$ with $n$ a (large) positive integer, and indicates that there emerges a corresponding three-body bound state whose binding energy $E_n$ scales as $\ln|E_n|\sim-(n\pi-\theta)^2/4$. The universal scaling of $E_n$ does not depend on the mass ratio $\gamma$ because in the current setup, we could rescale momentum for each atom independently [cf.~Eq.~(\ref{rrs})], which shall not change the low energy (large $b$) behavior of Eq.~(\ref{v3rg}). As we shall see below, for systems with non-zero co-dimensions, this independence shall no longer hold.

\emph{Three-body bound states.} 
To justify the conclusion by the above perturbative RG analysis, we solve explicitly the three-body bound states of the corresponding single-channel model, in which the interaction Hamiltonian is $H_{\rm int}=\lambda\psi^\dagger_A(\mathbf{0})\psi^\dagger_B(\mathbf{0})\psi_B(\mathbf{0})\psi_A(\mathbf{0})$. Employing the path integral representation and after the Hubbard-Stratonovich transformation, we find the system Lagrangian
\begin{align}
L=&\sum_{j=A,B}\int d^2\boldsymbol\rho \psi^\dagger_j(\boldsymbol\rho)\left(i\partial_t+\frac{\nabla^2}{2 m_j}\right)\psi_j(\boldsymbol\rho)-\frac{1}{\lambda}a^\dagger a\notag\\
&-[a^\dagger\psi_A(\mathbf{0})\psi_B(\mathbf{0})+\text{H.c.}]\label{bound},
\end{align}
where $a$ is the auxiliary field and an implicit momentum cutoff $\Lambda_0$ is imposed. Within this model (\ref{bound}), we tune $\lambda$ to make the two-body scattering on resonance.

\begin{figure}
	\includegraphics[width=0.47\textwidth]{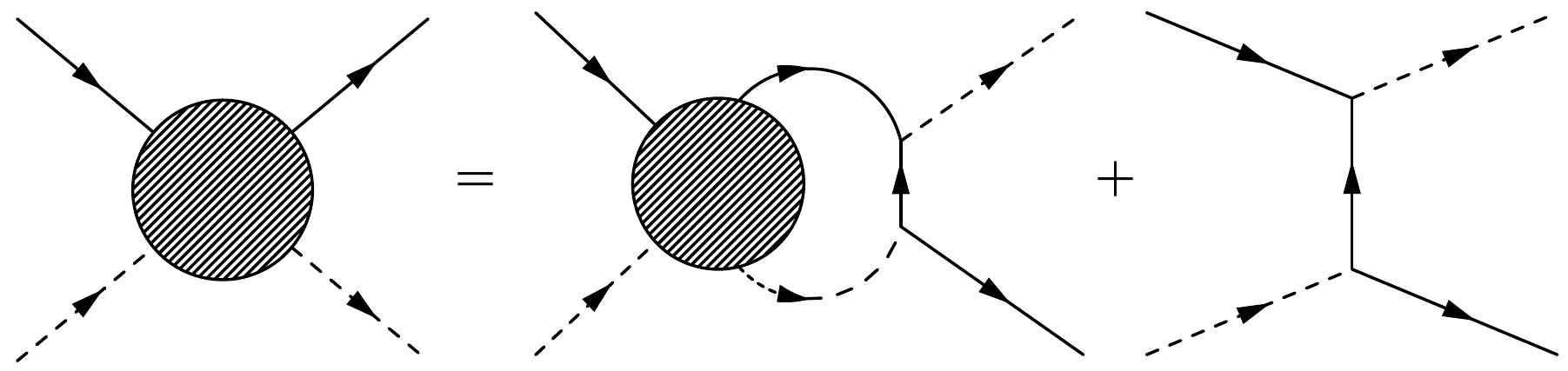}
	\caption{The STM equation for the atom-dimer scattering. The dashed lines represent the dimer field and the solid lines represent the atom fields.}
	\label{STM}
\end{figure}

The properties of the three-body bound states can be determined by studying the full scattering amplitude between an atom and dimer of Eq.~(\ref{bound}). The full scattering amplitude $B(\mathbf{k}_1,\mathbf{k}_2, E)$ with
$\mathbf{k}_1 (\mathbf{k}_2)$ the momentum for the incoming (outgoing) $A$ atom and $E$ the on shell energy satisfies the Skornyakov-Ter-Martirosyan (STM) equation represented in FIG.~\ref{STM}. 
In the limit that $E$ approaches the binding energy $E_B$ of a certain three-body bound state whose internal wave-function is $\phi(\mathbf k)$, $B(\mathbf{k}_1,\mathbf{k}_2,E)=\phi(\mathbf k_1)\phi^*(\mathbf k_2)/(E-E_B)$, and in the $s$-wave channel the STM equation represented in FIG.~\ref{STM} is transcribed into
\cite{Zhang2017}
\begin{align}
\phi(k)=&\int^{\Lambda_0}_0\frac{qdq}{(2\pi)^2}\ln\left(\frac{k^2+q^2-2Em_A}{\Lambda_0^2m_A/m_B}\right)\notag\\&\frac{{2\pi^2\phi(q)}}{\left(q^2/2-m_AE\right)\ln\left(\sqrt{q^2/2-Em_A}/\Lambda_0\right)}.\label{wfeq}
\end{align}
From Eq.~(\ref{wfeq}), it is clear that if we define $\tilde E\equiv m_A E$, the mass ratio $m_A/m_B$ only appears in $\ln(\Lambda_0^2 m_A/m_B)$, whose dependence is negligible for large cutoff $\Lambda_0$. The asymptotic behavior of the zero energy wave-function could be obtained by the leading logarithm approximation \cite{Son1999}. Within this approximation, we differentiate both sides of Eq.~(\ref{wfeq}) and find that the wave-function satisfies a simplified equation
\begin{align}
\phi''(z)=-2\phi(z)\label{swfeq}
\end{align}
where $z\equiv-\ln(k/\Lambda_0)$. Equation~(\ref{swfeq}) amounts to the Sch\"odinger equation in the momentum space.
The solution to Eq.~(\ref{swfeq}) has the general form $\phi(z)=\sqrt{z}[\cot(\theta-3\pi/4)J_1(2\sqrt{2z})+Y_1(2\sqrt{2z})]$ with $\theta$ the three-body parameter brought about by imposing a boundary condition on $\phi(z)$ at momentum $k^*$ which is much smaller than $\Lambda_0$. Therefore in the regime $0<k<k^*$, there are a infinite number of zero points of $\phi(z)$ approximately whereas $2\sqrt{2\ln(\Lambda_0/k)}=n\pi-\theta$. Compared with the zero energy wave-function of the Efimov states in the real space \cite{Braaten2006}, these zero points of $\phi(z)$ indicate the existence of bound states whose energies are $E_n\sim-\exp[-(n\pi-\theta)^2/4]$ for large $n$.

We have calculated the binding energies by solving Eq.~\eqref{wfeq} numerically. The result for $m_A=m_B$ is given in TAB.~\ref{numeric}. We find that the scaling of the binding energies of the shallower bound states approaches the one derived by the analytic methods above. We have also checked that the mass ratio does not affect the scaling behavior.
\begin{table}[t]
	\centering 
	\begin{tabular}{c|c|c}
	\hline\hline
  $n$ & $\varphi_n$  & $\varphi_n-\varphi_{n-1}$\\ 
 \hline
 1  & 2.3315 &   \\ 
		2 & 3.3780 &1.0465 \\ 
		3  & 4.4010 &1.0230   \\ 
		4  & 5.4150&1.0140  \\ 
		5  & 6.4245&1.0095 \\ 
  \hline
  	\end{tabular}
	
	\caption{The numerical result of the three-body binding energies $E_n$ of Eq.~(\ref{wfeq}) for $m_A=m_B$. Here $2m_AE_n/\Lambda_0^2=-\exp(-\pi^2\varphi_n^2/4)$. The difference $\varphi_n-\varphi_{n-1}\to1$ when the bound states are becoming shallower. }\label{numeric}
\end{table}

\emph{Generalized RG analysis.} 
The perturbative RG analysis of Eq.~(\ref{model}) has led to the same conclusion on the three-body bound states as by the explicit calculation based on the STM equation. Similar analysis can guide us to search universal three-body bound states beyond the Efimov paradigm in other mixed dimension setups on two-body interaction resonances. We consider a generalized model of two bosonic $A$ atoms and a single $B$ atom as
\begin{align}
L=&\sum_{j=A,B}\int d^{d_j}x \psi^\dagger_j\left(i\partial_t+\frac{\nabla^2}{2 m_j}\right)\psi_j\notag\\&+\int d^{d_{\rm int}}x\left[D^\dagger\left(i\partial_t+\frac{\nabla^2}{2M}-\bar\nu\right)D-\bar v_3\psi_A^\dagger D^\dagger D \psi_A\right]\notag\\
&-\bar g\sum_{\mathbf{q,k,k_A,k_B}}\left[q^\ell f_\ell(\hat{\Omega}_\mathbf q)\tilde D^\dagger(\mathbf{k})\tilde\psi_A\left(\mathbf{\frac{k}{2}+q},\mathbf{k}_A\right)\right.\notag\\
&\left.\times\tilde\psi_B\left(\mathbf{\frac{k}{2}-q},\mathbf{k}_B\right)+\text{H.c.}\right],\label{general}
\end{align}
where the $A$ ($B$) atoms move in a $d_A$($d_B$)-dimensional space, and the dimer $D$ moves in the $d_{\rm int}$-co-dimensional space in which the $A$ and $B$ atoms interact in $\ell$-wave. $\tilde D$ and $\tilde\psi_j$ are the Fourier transform of $D$ and $\psi_j$, and $f_\ell(\hat{\Omega}_\mathbf q)$ is an angle function. There is a physical constraint $d_{\rm int}\leq d_A(d_B)$. For the sake of physical realization, we take $d_j\le3$. 
In the case $d_{\rm int}=0$, $\ell$ has to be zero and $f_0=1$. When $d_{\rm int}=1$, there are two possibilities: $\ell=0$ and $f_0=1$, and $\ell=1$ and $f_1(\hat{\Omega}_\mathbf q)=\text{sgn}(q)$. 
For $d_{\rm int}=2$, $f_\ell(\hat{\Omega}_\mathbf q)=\exp(i\ell\phi_q)$, and for $d_{\rm int}=3$, 
$f_\ell(\hat{\Omega}_\mathbf q)=Y_{\ell m}(\theta_q,\phi_q)$. 

The RG equation \eqref{v3rg} derived from Eq.~(\ref{model}) predicts the existence of the universal three-body bound states whose binding energies scale as $\ln|E_n|\sim-(n\pi-\theta)^2/4$. There are two key ingredients in Eq.~\eqref{v3rg}: (I) the presence of $1/b$ on the right hand side of the equation, which derives from the logarithmic dependence of the two-body phase shift expansion $\cot\delta(k)=-1/Sk^2+2\log(k R)/\pi$. (II) the three-body interaction is marginal. 
Thus we require the general model (\ref{general}) to have the same properties for the two-body scattering and the three-body interaction. We apply power counting on the diagram in FIG. \ref{RG} (a) which yields the two-body phase shift and find $d_A+d_B-d_{\rm int}+2\ell=4$ as the condition for the phase shift to retain the structure $\cot\delta(k)=-1/Sk^2+2\log(k R)/\pi$. Since the dimension of $\bar g$ is zero [cf.~Eq.~(\ref{g})], a direct power counting analysis of Eq.~(\ref{general}) yields $d_B-d_{\rm int}+2\ell=2$ as the condition for the three-body interaction being marginal, i.e., the dimension of $\bar v_3$ is zero. 

\begin{table}
	\centering 
	\begin{tabular}{m{0.8cm} m{0.8cm}  m{0.8 cm} m{0.8 cm} m{0.8 cm} m{4 cm} }
		\hline\hline
	  Case &  $d_A$ & $d_B$ & $d_{\rm int}$ & $\ell$& $\ln |E_n|$ \\ 
		\hline
		(I) & $2$  & $2$ &   $0$ & $0$ & $-(n\pi-\theta)^2/4$\\ 
		(II) & $2$ & $3$ & $1$ & $0$ & $-\frac{(n\pi-\theta)^2}{4}\frac{\sqrt{\gamma(2+\gamma)}}{1+\gamma}$\\ 
		(III) &$2$  & $2$&$2$&$1$& $-2\exp\left[\frac{n\pi \gamma(2+\gamma)}{1+\gamma}+\theta\right]$ \\ 
		(IV)&$2$  & $1$ & $1$ & $1$ & $-2\exp\left[n\pi \sqrt{h(\gamma)/2}+\theta\right]$\\
		\hline
	\end{tabular}
	
	\caption{Universal scaling of binding energies $E_n$ of three-body bound states in mixed dimensions with $\ell$-wave two-body resonant scattering. Here $\theta$ is a three-body parameter and the scaling function $h(\gamma)$ is given in Eq.~(\ref{h}).}\label{four}
\end{table}

These two conditions are satisfied when $d_A=2$ and $d_B=d_{\rm int}+2-2\ell$. Correspondingly $\ell$ is either zero or $1$. We list all four possibilities in TAB.~\ref{four}.
We have studied previously Case (I) in detail. For Case (II), based on the diagram in FIG. \ref{STM}, 
we proceed to work out the RG equation for $\bar v_3$ which is similar to Eq.~(\ref{v3rg}), and predict that there shall be universal three-body bound states whose binding energies scale as $\ln |E_n|\sim-\frac{(n\pi-\theta)^2}{4}\frac{\sqrt{\gamma(2+\gamma)}}{1+\gamma}$. Here the mass ratio dependence of the scaling comes from the fact that the co-dimension is nonzero. The smaller the mass ratio $\gamma$ is, the denser the bound states are in the energy domain. If we choose $^{133}$Cs as the $A$ atoms and $^6$Li as the $B$ atoms, between which there is an $s$-wave Feshbach resonance at the magnetic field $842.75$ G, the factor ${\sqrt{\gamma(2+\gamma)}}/({1+\gamma})$ can be suppressed to $0.29$. 
For the last two cases, since $d_B=d_{\rm int}$ and there is no longer a momentum integration for the diagram in FIG. \ref{RG} (c), one needs to go further to include the one-loop diagrams in the RG equation for $\bar v_3$ \cite{Nishida2013,Nishida2014}. Case (III) is the super Efimov states which have been discovered before \cite{Nishida2013, Nishida2014, Zinner2014, Gridnev2014, Gao2015}. We find that Case (IV) holds an infinite series of universal three-body
bound states whose binding energies also following the super Efimov scaling, i.e., $\ln|E_n|\sim-2\exp\left[n\pi \sqrt{h(\gamma)/2}+\theta\right]$ with the scaling function
\begin{align}
h(\gamma)=\frac{\gamma\sqrt{2+\gamma}}{(1+\gamma)\left[\sqrt{\gamma}+4\sqrt{\gamma^3}+2\sqrt{\gamma^5}-2\gamma(1+\gamma)\sqrt{2+\gamma}\right]}.\label{h}
\end{align}
Note that there may exist other possible infinite series of universal three-body bound states in mixed dimensions beyond our above RG analysis. We defer the problem of two fermionic $A$ atoms interacting with a single $B$ atom in mixed dimensions to a future study.

\emph{Experimental detection.} Mixed dimensions have prospective realizations in cold atom systems in which atoms can be tailored to move in a space of dimension $d<3$ by applying external confinements \cite{Bloch2008}. Experimental signatures of the super Efimov effect have been studied in Refs.~\cite{Gao2015, Zhang2017}. Here we focus on Case (I) and (II) listed in TAB.~\ref{four}. The two-body interactions between the $A$ and $B$ atoms restricted in co-dimension $d_{\rm int}=0,1$ can be achieved by an optically controlled magnetic Feshbach resonance \cite{Chin2015}. To realize Case (I), one can start with confining the $A$ and $B$ atoms in a two-dimensional plane and tunes the external magnetic field in the vicinity of an $\ell$-wave magnetic Feshbach resonance such that the background scattering is negligible. Furthermore, one intersects the plane with a well focused laser beam, which gives rise to vector light shifts of the energies of the atoms and the Feshbach molecules \cite{Spielman2014}. By controlling the frequency and intensity of the laser, one can push the scattering between an $A$ and $B$ atom sitting right at the Feshbach resonance when the two atoms simultaneously enter into the cross section of the light beam in the plane \cite{Chin2015}. The linear dimension of the laser cross section, which can be as small as the order of the laser wavelength, serves as the short distance cutoff of our model (\ref{model}). A similar scheme can be set up for Case (II), in which resonant scattering in one-co-dimension can be realized within the laser beam which induces the vector light shifts. Further reduction in the short distance cutoff of atomic resonant scattering in co-dimensions may make use of type-II superconductors where a magnetic flux is pinned through a vortex core and confined within a cross section of size down to tens of nanometers \cite{Tinkham}.

The effects of the predicted universal three-body bound states in Case (I) and (II) could be detected in observables such as the time of flight and radio-frequency association spectrum. The Efimov effect was predicted to give rise to a tail in the atom momentum distribution $n_k\sim \sin[2s_0\ln(k/\kappa)]/k^5$ by the field theoretic method of the operator product expansion (OPE). Here $s_0\approx1.006$ and $\kappa$ is the Efimov three-body parameter. The magnitude of the tail is proportional to the three-body contact \cite{Braaten2011}. Such a tail has been observed in the time of flight experiment of unitary Bose gases \cite{Jin2014, Braaten2014}. By the same OPE method, we find that the three-body bound states in Case (I) and (II) yield a momentum tail of the $A$ atoms $n_k\sim\cos[4\sqrt{-\ln(k R)/s}+2\theta]/[k^4 \sqrt{-2\ln (k R)}]$. In the situation $m_A=m_B$, we have $s=1$ for Case (I) and $s=(3/4)^{1/2}$ for Case (II). 
Another way to probe the predicted universal three-body bound states is via the radio-frequency association spectrum. 
Experimentalists have employed a radio-frequency field to associate three distinguishable fermionic atoms of $^6$Li into an Efimov trimer state \cite{jochim_radio, ueda_radio}. Scanning the frequency of the field, they can pin down the trimer binding energy by looking for a maximum in the association rate. The same scheme can be applied in our cases as well. 

\emph{Acknowledgements.}
We thank X. Cui for discussions. This research was supported in part by NSFC Grant No. 11474179. 

\emph{Note added.} Recently Ref.~\cite{Nishida2017} found universal four-boson bound states with energies scaling $\ln|E_n|\sim-2(n\pi+\theta)^2/27$ at three-body interaction resonance in two dimensions. Our mixed dimensional systems only require two-body interaction resonance, which eases their experimental realization.

\end{document}